\documentclass[prb,twocolumn,superscriptaddress]{revtex4-1}
\usepackage{graphicx}
\usepackage{hyperref}
\usepackage{amsmath}

\begin{document}
\title{Spin-torque-induced magnetization dynamics in ferrimagnets based on Landau-Lifshitz-Bloch Equation}

\author{Zhifeng Zhu}
\email{a0132576@u.nus.edu}
\affiliation{Department of Electrical and Computer Engineering, National University of Singapore,
	Singapore 117576 }
\author{Xuanyao Fong}
\affiliation{Department of Electrical and Computer Engineering, National University of Singapore,
	Singapore 117576 }
\author{Gengchiau Liang}
\email{elelg@nus.edu.sg}
\affiliation{Department of Electrical and Computer Engineering, National University of Singapore,
	Singapore 117576 }
\date{\today}

\begin{abstract}
A theoretical model based on the Landau-Lifshitz-Bloch equation is developed to study the spin-torque effect in ferrimagnets. Experimental findings, such as the temperature dependence, the peak in spin torque, and the angular-momentum compensation, can be well captured. In contrast to the ferromagnet system, the switching trajectory in ferrimagnets is found to be precession free. The two sublattices are not always collinear, which produces large exchange field affecting the magnetization dynamics. The study of material composition shows the existence of an oscillation region at intermediate current density, induced by the nondeterministic switching. Compared to the Landau-Lifshitz-Gilbert model, our developed model based on the Landau-Lifshitz-Bloch equation enables the systematic study of spin-torque effect and the evaluation of ferrimagnet-based devices.
\end{abstract}
\maketitle

\section{Introduction}
Ferrimagnets (FiMs) with antiferromagnetic exchange coupled transition-metal (TM) and rare-earth (RE) alloys have attracted considerable attention due to the rich physics \cite{PhysRevLett.97.217202,Radu2011,Ostler2012,PhysRevLett.118.167201,PhysRevB.96.100407,PhysRevB.95.140404,PhysRevLett.119.197201} and their promise in device applications \cite{doi:10.1063/1.4916665,Kim2017,PhysRevB.97.184410}. The FiMs are expected to have fast spin dynamics like antiferromagnets (AFMs), but their magnetic states can be electrically sensed using the tunnel magnetoresistance (TMR) effect due to the finite net magnetization ($\mathbf{m_{net}}$), which can be tuned by temperature ($T$) or material composition ($X$). In addition, the FiMs have large bulk perpendicular anisotropy, which offers an alternative to the ferromagnets (FMs) and enables the scaling down of MRAM down to 20 nm \cite{doi:10.1063/1.4916665}. Furthermore, different g factors between sublattices induce an angular-momentum compensation point, which enables fast domain-wall motion \cite{Kim2017}.

The FiMs can be manipulated by magnetic field or laser heating \cite{Radu2011,Ostler2012,PhysRevB.65.012413,PhysRevLett.99.047601,PhysRevB.85.104402,PhysRevLett.108.127205,Barker2013}, but an electrical method, such as the spin-transfer torque (STT) \cite{PhysRevLett.97.217202} or the spin-orbit torque (SOT) \cite{PhysRevLett.118.167201,doi:10.1063/1.4962812}, is preferred for electrical characterizations and applications. Therefore, it is important to study the magnetization dynamics under spin torque using a model which can incorporate the effects of $T$ and $X$. However, the commonly used theoretical model based on the Landau-Lifshitz-Gilbert (LLG) equation \cite{PhysRevLett.97.217202,PhysRevB.73.220402,PhysRevB.74.134404,OEZELT201528} is limited at fixed $T$ due to the assumption of a fixed magnetization length (see Appendix A for detailed analysis of the LLG model). In contrast, the Landau-Lifshitz-Bloch (LLB) model has been widely used to describe the magnetization dynamics at elevated $T$ \cite{PhysRevB.55.3050}, where the $T$-induced magnetization-length change is taken into account by including a longitudinal relaxation term. To date, the LLB equations have been implemented for both FMs \cite{PhysRevB.55.3050} and FiMs \cite{PhysRevB.86.104414}, and recently the effect of spin torque in FMs has also been included \cite{PhysRevB.80.094418}. Starting from the atomistic Landau-Lifshitz equation, in this work, we extend the LLB model to capture the spin-torque effect in FiMs. The numerical simulation of current-induced switching in a FiM/heavy-metal (HM) bilayer is then performed, and we find the modified LLB model can reproduce salient experimental findings, such as the magnetization compensation, the reversal of switching direction across the magnetization-compensation temperature ($T_{MC}$) \cite{1882-0786-9-7-073001}, and the peak in spin torque at $T_{MC}$ \cite{doi:10.1063/1.4985436}. In addition, the spin-torque-induced sublattice dynamics in FiMs is studied and compared to that in AFMs and FMs. The switching trajectory is found to be precession free, and the sublattices are not always collinear. Finally, the effect of $X$ on FiM properties is studied.
\section{THE MODIFIED LANDAU-LIFSHITZ-BLOCH EQUATION}
\begin{figure}
	\centering
	\includegraphics[width=\columnwidth]{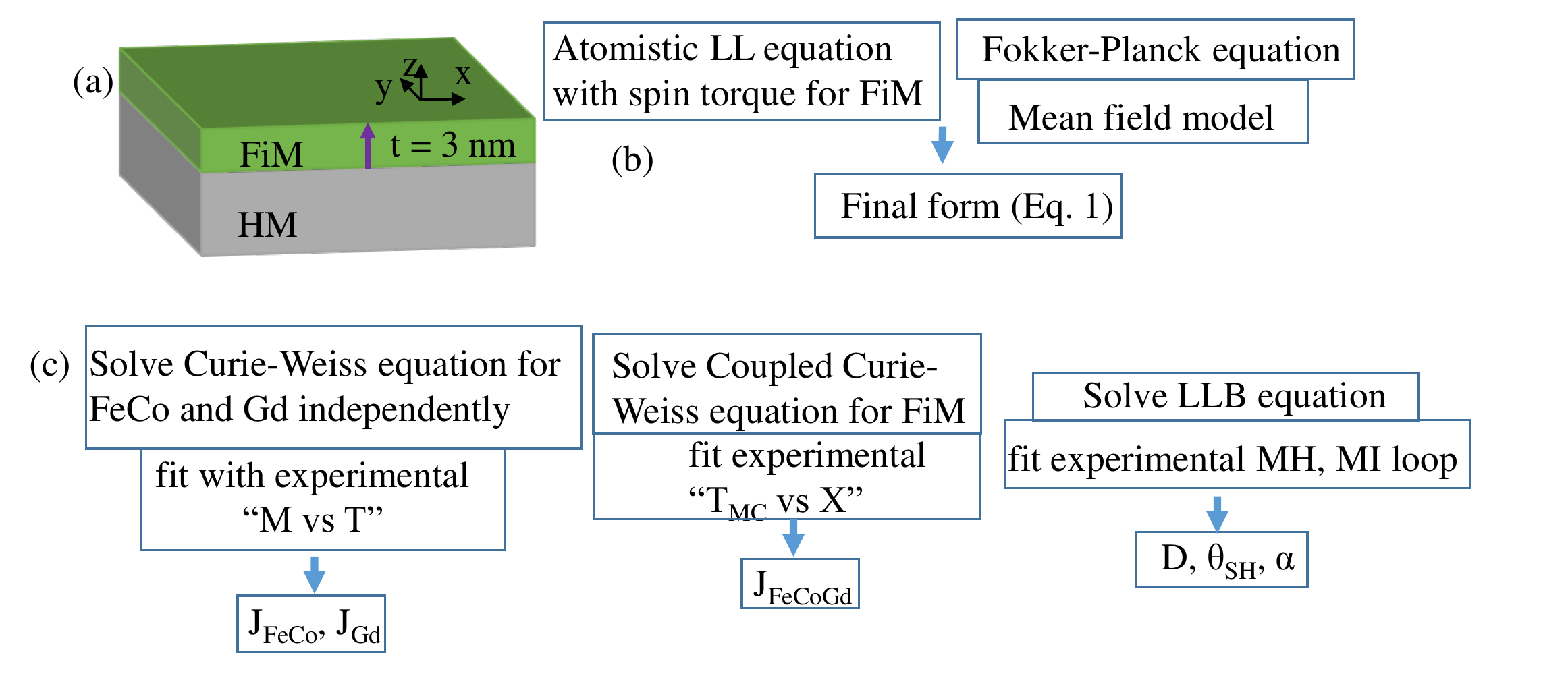}
	\caption{(a) Schematic view of the device structure consisting of a FiM layer deposited on top of a HM. The FiM in this study is perpendicularly magnetized Gd$_X$(FeCo)$_{1‒X}$ alloy, where Gd and FeCo are antiferromagnetically coupled. Procedure of (b) equation derivation, (c) model validation and parameter determination. The derivation starts from the Landau-Lifshitz equation including the spin torque, followed by the corresponding Fokker-Planck equation to account for the statistic behavior, and yields the final form after using the mean field approximation. This model is validated by comparing with the experimental “$M$ vs $T$ trajectory \cite{PhysRev.132.1092}”, “$T_{MC}$ vs Gd concentration ($X$)”, and “$M$-$H$ and $M$-I loops \cite{doi:10.1063/1.4962812}”.}
	\label{fig1}
\end{figure}

As shown in Fig.~\ref{fig1}(a), the device structure we studied consists of a FiM (Gd$_X$(FeCo)$_{1‒X}$) deposited on top of a HM layer. The magnetizations of sublattices are manipulated by the SOT generated by in-plane electrical current. The FiM is treated as a two-sublattice model, i.e., Gd and FeCo, which is justified by the experimental observation that the magnetizations of Fe and Co are parallel up to the Curie temperature ($T_C$) \cite{PhysRevB.84.024407}. The magnetization dynamics of each sublattice is captured by the modified LLB equation \cite{Radu2011,Ostler2012,PhysRevB.80.094418,PhysRevB.84.024407,doi:10.1002/9780470022184.hmm205,PhysRevB.86.214416,doi:10.1142/S201032471650003X,PhysRevB.74.094436,0953-8984-26-10-103202} (see Appendix B for the derivation)
\begin{equation}\label{eq1}
\begin{aligned}
\dot{\mathbf{m}}_v
&=\gamma_v(\mathbf{m}_v{\times} \mathbf{H}_{v}^{MFA})-\Gamma_{v,\parallel}(1-\frac{(\mathbf{m}_v\cdot\mathbf{m}_{0,v})}{m_v^2})\mathbf{m}_v \\
&-\Gamma_{v\perp}\frac{\mathbf{m}_v{\times}(\mathbf{m}_v{\times}\mathbf{m}_{0,v})}{m_v^2},\\
\end{aligned}
\end{equation}
where $\Gamma_{v,\parallel}=\Lambda_{v,N}B(\xi_{0,v})/(\xi_{0,v}B'(\xi_{0,v}))$ and $\Gamma_{v,\perp}=\Lambda_{v,N}[\xi_{0,v}/B(\xi_{0,v})-1]/2$ are the coefficients of longitudinal and transverse relaxation, respectively. The dimensionless field is given by 
\begin{equation}\label{eq2}
\mathbf{\xi}_{0,v}=\beta\mu_v(\mathbf{H}_v^{MFA}+\mathbf{H}_I/\lambda_v).
\end{equation}
Eq.~(\ref{eq1}) contains two coupled equations for FeCo and Gd identified by the subscript $v$, which need to be solved simultaneously \cite{PhysRevB.97.184410}. The first term on the right hand side describes the magnetization precession around the mean-free field 
\begin{equation}\label{eq3}
\mathbf{H}_{v}^{MFA}=\mathbf{H}_{ext}+\mathbf{H}_{A,v}+(J_{0,v}/\mu_{v})\mathbf{m}_{v}+(J_{0,vk}/\mu_{v})\mathbf{m}_{k},
\end{equation}
which consists of the external magnetic field $\mathbf{H}_{ext}$, the crystalline anisotropy field $\mathbf{H}_{A,v}=(2D_v/\mu_v)m_{v,z}\mathbf{e}_z$ with coefficient $D_v$, and the exchange coupling between sublattices with coefficients $J_{0,v}$ and $J_{0,vk}$. The damping coefficient is given by
\begin{equation}\label{eq4}
\Lambda_{v,N}=2\gamma_v\lambda_v/(\beta\mu_v),
\end{equation}
\begin{equation}\label{eq5}
\beta=1/(k_BT),
\end{equation}
where $\lambda_v$ is the damping constant, $\gamma_v$ is the gyromagnetic ratio, $\mu_v$ is the magnetic moment, and $k_B$ is the Boltzmann constant. As previously mentioned, the $T$-induced magnetization-length change is described by the longitudinal relaxation term using the Brillouin function 
\begin{equation}\label{eq6}
B(\xi)=coth(\xi)-1/\xi,
\end{equation}
and the spin-torque effective field $\mathbf{H}_I$ is given by 
\begin{equation}\label{eq7}
\mathbf{H}_I=\mathbf{J}_S\hbar/(2et_{FiM}|XM_{S,v}-qM_{S,k}|),
\end{equation}
where $\hbar$ is the reduced Planck constant, $e$ is the electron charge, $t_{FiM}$ is the thickness of FiM layer, and $M_S$ is the saturation magnetization. The spin current density $J_S$ is formulated as 
\begin{equation}\label{eq8}
\mathbf{J}_S={\theta}_{SH}\mathbf{\sigma}\times\mathbf{J}_C,
\end{equation}
where $\theta_{SH}$ is the spin-Hall angle, $\mathbf{\sigma}$ is the polarization of spin current, and $\mathbf{J}_C$ is the charge current. The equilibrium magnetization $\mathbf{m}_{0,v}$ is calculated via the coupled Curie-Weiss equation
\begin{equation}\label{eq9}
\mathbf{m}_{0,v}=B({\xi}_{0,v})\mathbf{\xi}_{0,v}/\xi_{0,v}.
\end{equation}
Similar to the LLB in FM \cite{PhysRevB.80.094418}, the effect of spin torque only enters the two relaxation terms. Furthermore, Eq.~(\ref{eq1}) reduces to Eq. (4) in Ref. \cite{PhysRevB.86.104414} when the spin torque vanishes, or to Eq. (7) in Ref. \cite{PhysRevB.80.094418} when FeCo and Gd are not distinguished. The numerical integration of Eq.~(\ref{eq1}) proceeds using a fourth-order predictor-corrector method \cite{PhysRevB.97.184410}.

The parameters used in the simulation are determined as follows [see Fig.~\ref{fig1}(c)]: First, the Curie-Weiss equation [Eq.~(\ref{eq9})] for pure Gd and FeCo is solved independently and fit to the experimental $M$-$T$ curves \cite{PhysRev.132.1092} to determine the exchange coupling coefficients $J_{Gd}=0.98\times10^{-21}$J and $J_{FeCo}=1.5\times10^{-21}$J. Then, the $J_{GdFeCo}=-7.63\times10^{-21}$J is obtained by solving the coupled Curie-Weiss equations of FiM. In addition, a sufficient anisotropy is used to ensure the perpendicular magnetization. The $\theta_{SH}$ and $\lambda$ are swept with $D$ to fit the experimental $M$-$H$ and $M$-$J$ loops \cite{doi:10.1063/1.4962812}, and a good agreement \cite{PhysRevB.97.184410} with the experimental data is obtained with $\theta_{SH}=0.0037$\cite{PhysRevB.83.174405}, $\lambda$=0.07, and $D=3.2\times10^{-26}$J. 

\section{DETERMINISTIC SWITCHING INDUCED BY THE SPIN-ORBIT TORQUE}
\begin{figure}
	\centering
	\includegraphics[width=\columnwidth]{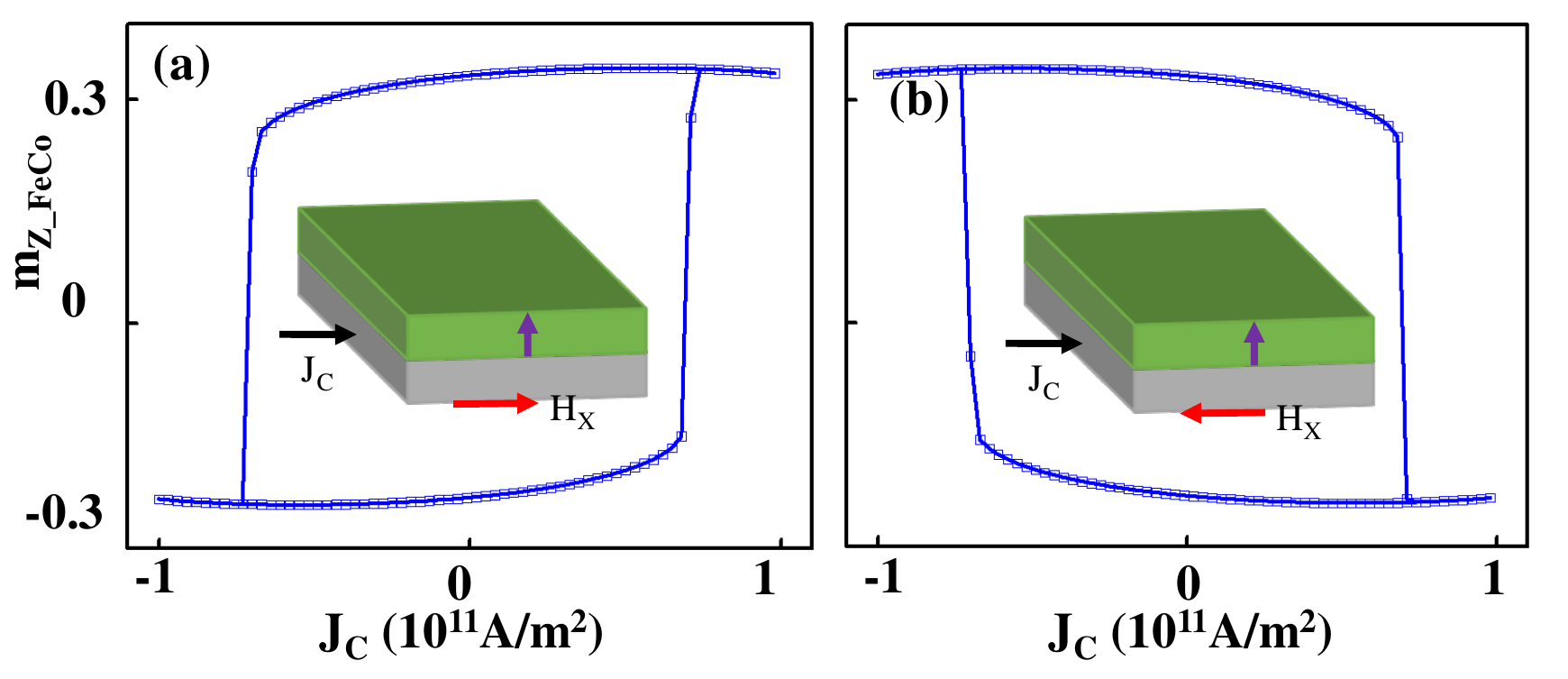}
	\caption{Simulated SOT-induced switching in FiM Gd$_{21}$(FeCo)$_{79}$ under (a) $\mathbf{H}_X$ = 1 mT, (b) $\mathbf{H}_X=-1$ mT at $T$ = 300 K. The $\mathbf{H}_X$ only breaks the symmetry for deterministic switching. The reversal of switching direction under opposite $\mathbf{H}_X$ is similarly explained using the theory in perpendicular FM.}
	\label{fig2}
\end{figure}

We first study the SOT-induced deterministic switching in FiM using the LLB model. As shown in Fig.~\ref{fig2}, the $\mathbf{J}_C$ applied along the $x$ direction generates spin torque acting on the FiM layer due to the spin-hall effect (SHE) or the inverse spin galvanic effect (ISGE) \cite{MihaiMiron2010,PhysRevB.79.094422,Kurebayashi2014}. However, the magnetization cannot be switched vertically since the spin torque aligns the magnetization to $y$ axis. This is similar to the perpendicular FM switched by in-plane current, where an external field along the current direction ($\mathbf{H}_X$) is required to achieve deterministic switching \cite{Miron2011,PhysRevLett.109.096602,doi:10.1063/1.4902443,Fukami2016}. The switching in FM system can be understood as follows: The switching direction is determined by $\mathbf{H}_X$ as $\mathbf{L}={\Delta}\mathbf{m}\times\mathbf{H}_X$, and the spin torque, ${\Delta}\mathbf{m}=\mathbf{m}\times(\mathbf{m}\times\mathbf{H}_I)$, should be sufficient to overcome the energy barrier. Therefore, the switching direction will be reversed by reversing either $\mathbf{H}_X$ or current direction \cite{PhysRevLett.109.096602}. Recently, by applying $\mathbf{H}_X$, the current-induced deterministic switching in the FiM/HM bilayer has also been demonstrated \cite{PhysRevLett.118.167201,doi:10.1063/1.4962812}. The measured $M$-$J$ loop clearly shows an opposite switching direction by reversing the current, whereas the effect of $\mathbf{H}_X$ has not been investigated. In this study, we show that the switching direction is also reversed under opposite $\mathbf{H}_X$ [see Fig.~\ref{fig2}], which can be explained using the abovementioned two-torque analysis together with the exchange coupling between sublattices. As shown in Fig.~\ref{fig2}(a), the FiM at $T$ = 300 K is FeCo dominant. The positive $\mathbf{J}_C$ and $\mathbf{H}_X$ switch $\mathbf{m}_{FeCo}$ from down to up, and concurrently, the exchange interaction turns $\mathbf{m}_{Gd}$ from up to down. When the $\mathbf{H}_X$ is reversed, $\mathbf{m}_{FeCo}$ is switched from up to down [see Fig.~\ref{fig2}(b)], resulting in an opposite $M$-$J$ trajectory. To confirm the unique role of $\mathbf{H}_X$, we have verified that the equilibrium magnetization is not altered when only $\mathbf{H}_X$ is applied, and no switching event is observed when the current is swept with $\mathbf{H}_{ext}$ = 0 or $\mathbf{H}_Y$. Therefore, the SOT-induced switching in FiM is determined by the dominant sublattice, followed by the reversal of the other sublattice via exchange interaction, and the $\mathbf{H}_X$ only breaks switching symmetry. It is also worth noting that the maximum $m_Z$ in Fig.~\ref{fig2} is around 0.3, which is an evidence of the $T$-induced magnetization-length reduction with $m_Z$  = 1 defined at $T$ = 0 K.

\begin{figure}
	\centering
	\includegraphics[width=\columnwidth]{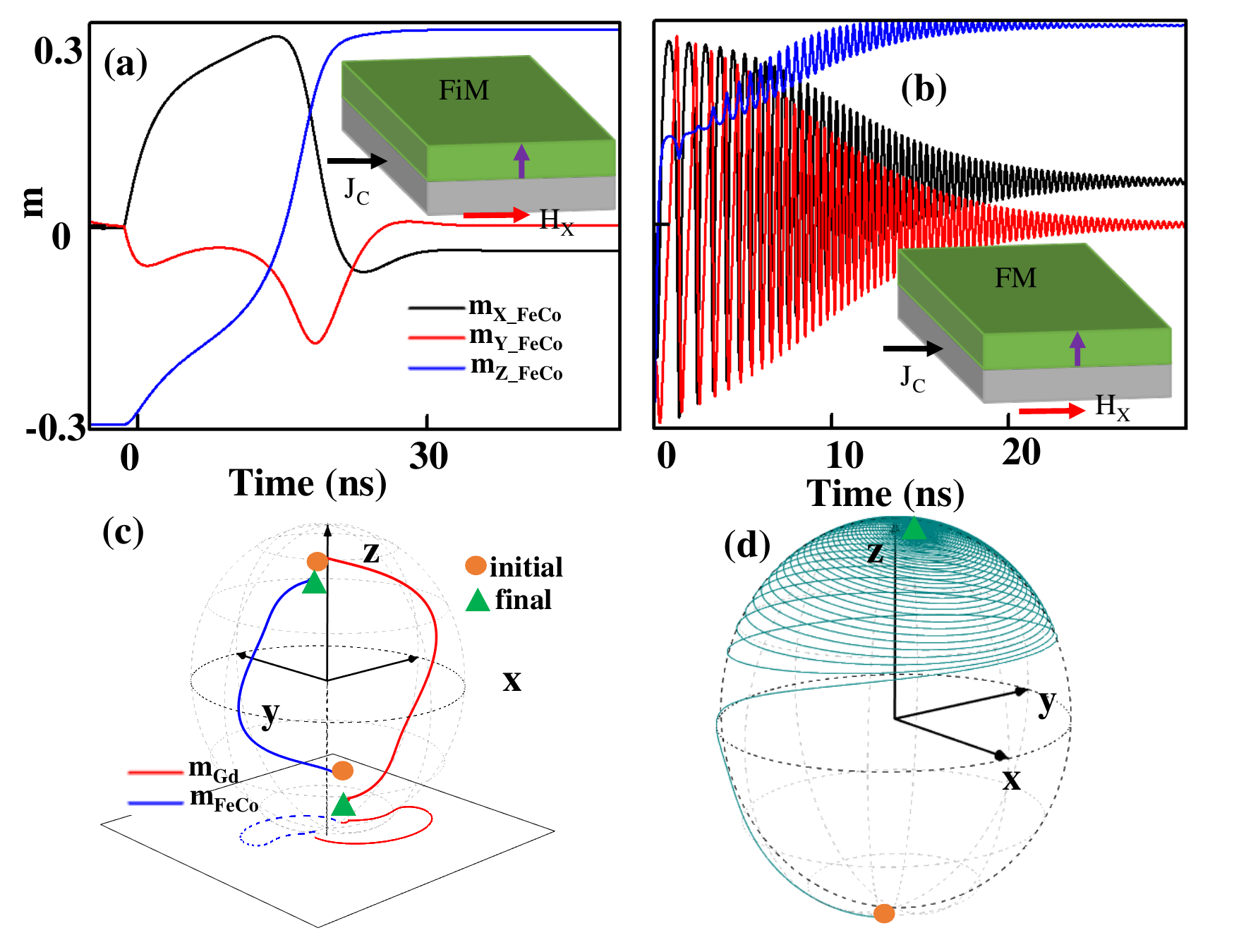}
	\caption{Time evolution of the SOT-induced switching (a) in FiM Gd$_{21}$(FeCo)$_{79}$ and (b) in perpendicular FM at $T$ = 300 K. (c) and (d) are the 3D trajectories corresponding to (a) and (b) respectively. The dot lines in (c) are the projections on the $x$-$y$ plane. All simulations start from the equilibrium state where $m_Z$ = 0.3. This reduced value reflects the $T$-dependent magnetization, where $m_Z$ = 1 is defined at $T$ = 0 K.}
	\label{fig3}
\end{figure}

\begin{figure}
	\centering
	\includegraphics[width=\columnwidth]{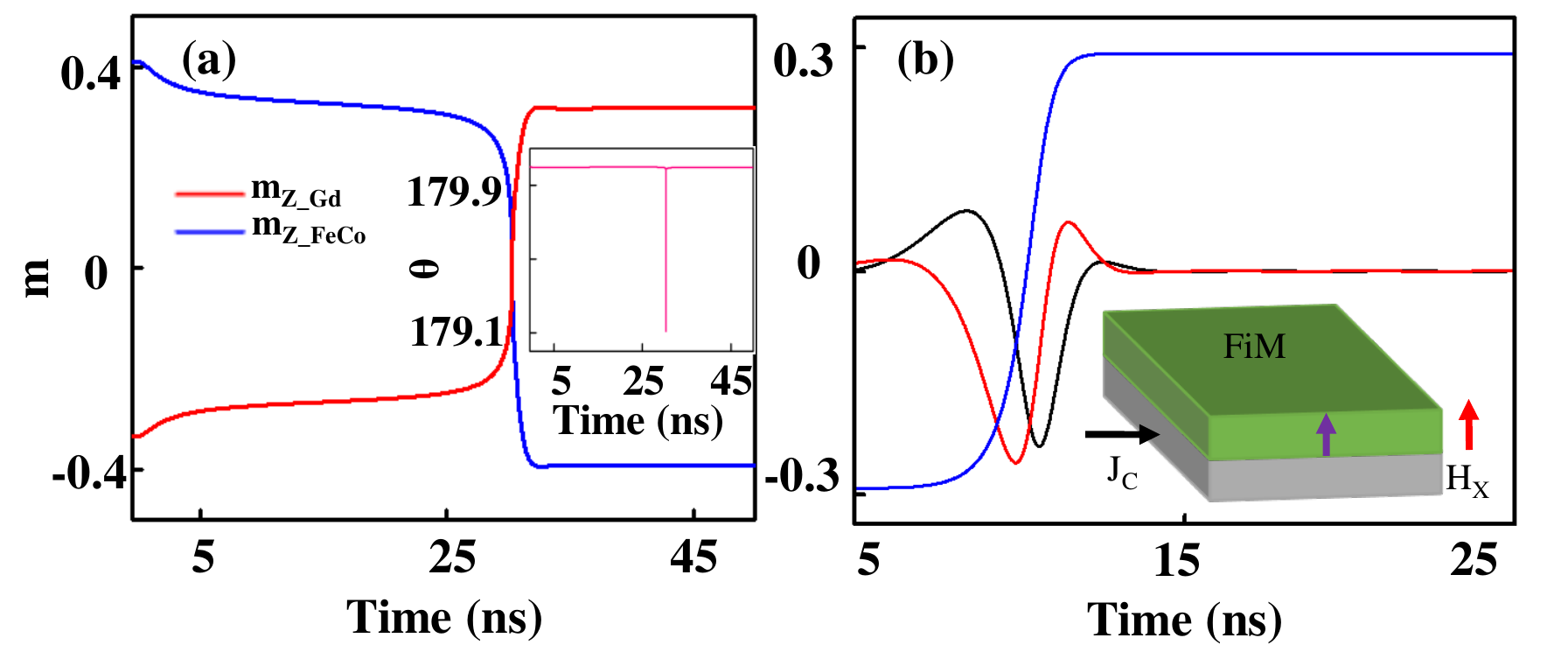}
	\caption{(a) Time evolution of $m_Z$ for FeCo and Gd sublattices at $T$ = 300 K, with inset showing the magnetization angle between $\mathbf{m}_{Gd}$ and $\mathbf{m}_{FeCo}$. (b) Field-induced switching in Gd$_{21}$(FeCo)$_{79}$ at $T$ = 300 K, which has similar trajectories with the current-induced switching.}
	\label{fig4}
\end{figure}

Although SOT and $\mathbf{H}_X$ have similar effects in switching FiM and perpendicular FM, the time evolutions of magnetization are very different as shown in Fig.~\ref{fig3}, i.e., the switching trajectory of FiM is precession free, whereas it is precessional in FM. In the SOT-switched FM with initial $m_Z= –1$, both anisotropy field and spin torque align the magnetization to the $+z$ direction for $m_Z>0$, resulting in a larger precession term compared to $m_Z < 0$, where the anisotropy field and spin torque are opposite. Consequently, more precession occurs when $m_Z> 0$  [see Fig.~\ref{fig3}(b)]. Similarly, the precession-free trajectory in FiM is attributed to the small precession term. As illustrated using the 3D trajectories in Fig.~\ref{fig3}(c), $\mathbf{m}_{Gd}$ and $\mathbf{m}_{FeCo}$ are switched to opposite directions. Due to the strong exchange coupling, many studies assume they are always collinear. However, as the time evolution of each sublattice and their relative angle shown in Fig.~\ref{fig4}(a), a maximum deviation of 0.9 degree is observed at $t$ = 30 ns. This number is similar to a recent report from Mishra et al. \cite{PhysRevLett.118.167201}, where a cant of one degree is estimated from the strength of exchange field. Since the exchange coupling between sublattices is very strong ($>$ 100 T \cite{PhysRev.82.565,Khymyn2017}), even a very small cant deviates the behavior of FiM from FM, which might contribute to the different magnetization dynamics shown in Figs.~\ref{fig3}(c) and ~\ref{fig3}(d). Similar noncollinearity between sublattices is also predicted in AFM \cite{doi:10.1063/1.4862467}, with the deviation angle determined by the strength of spin torque. To achieve a large-angle noncollinearity in FiM, recent study shows that a magnetic field over 5 T is required \cite{PhysRevLett.118.117203}. By studying the field-induced switching in FiM [see Fig.~\ref{fig4}(b)], a similar trajectory is observed compared to Fig.~\ref{fig3}(a), indicating that a large spin-torque effective field would be required to get a large angle deviation. However, as discussed in the next section, large spin torque aligns the magnetization to the spin direction, hence no switching happens. 

\section{EFFECT OF TEMPERATURE AND MATERIAL COMPOSITION}
\begin{figure}
	\centering
	\includegraphics[width=\columnwidth]{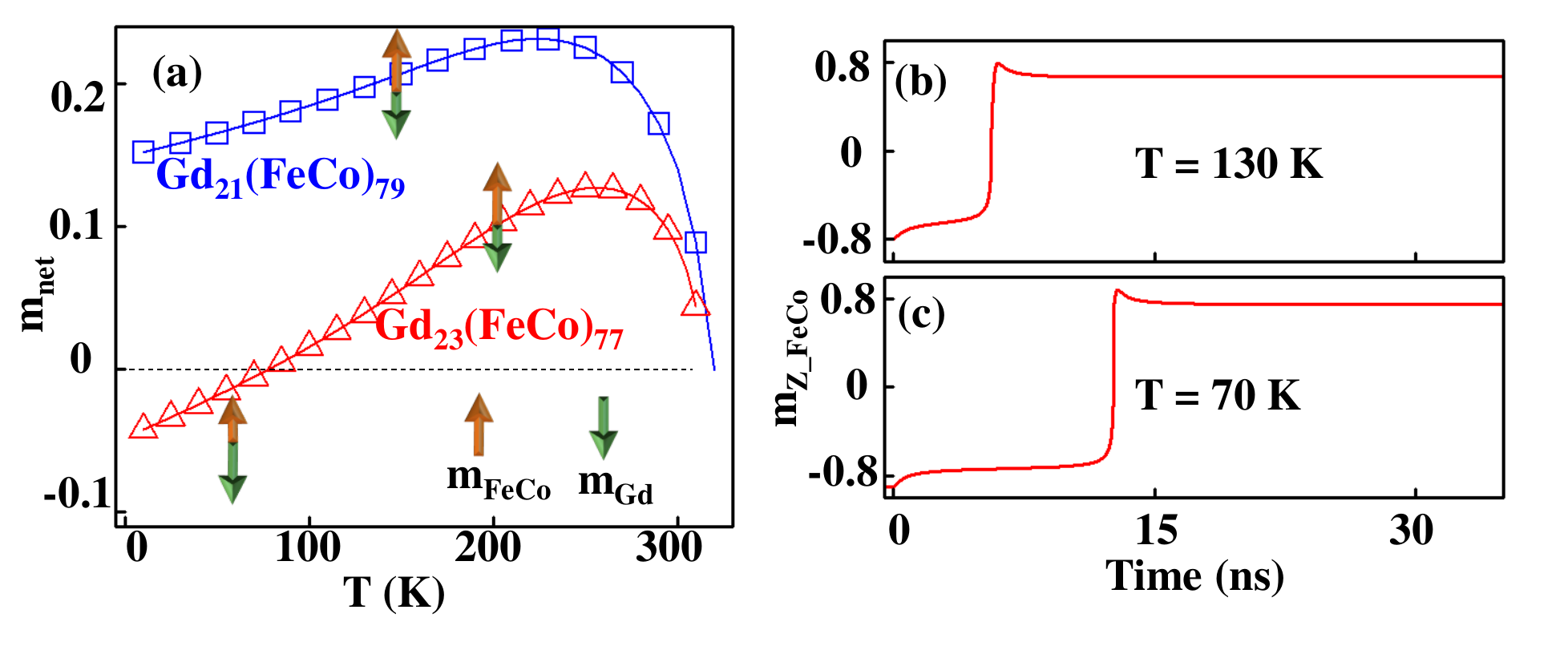}
	\caption{(a) Effect of $T$ on the net magnetization of Gd$_{21}$(FeCo)$_{79}$ (blue square) and Gd$_{23}$(FeCo)$_{77}$ (red triangle) with $T_{MC}$ = 75 K below which Gd is dominant, where the net magnetization is calculated using $m_{net}=(1-X)m_{FeCo}\mu_{FeCo}+Xm_{Gd}\mu_{Gd}$ with $\mu_{FeCo}=2.217\mu_B$ and $\mu_{Gd}=7.63\mu_B$. Time evolution of $m_{Z,FeCo}$ under spin torque at (b) $T$ = 130 K and (c) 70 K. }
	\label{fig5}
\end{figure}

$T$ and $X$ are often tuned in experiments to control the properties of FiM \cite{PhysRevLett.118.167201,Kim2017,1882-0786-9-7-073001,PhysRevB.96.064406}. By measuring the $M$-$H$ loops as a function of $T$, $T_{MC}$ can be identified where the coercive field ($H_C$) diverges. However, $T_{MC}$ may not exist in another sample with a different $X$ \cite{doi:10.1063/1.4962812}. In this study, the LLB equation is used to investigate two samples, i.e., Gd$_{21}$(FeCo)$_{79}$ and Gd$_{23}$(FeCo)$_{77}$, and we show that the existence of $T_{MC}$ is determined by the demagnetization speed and the relative magnitude of $\mathbf{m}_{FeCo}$ and $\mathbf{m}_{Gd}$. As reported in our recent study \cite{PhysRevB.97.184410}, both $\mathbf{m}_{FeCo}$ and $\mathbf{m}_{Gd}$ decrease with $T$ and vanish at the same temperature located between $T_{C,FeCo}$ (1043 K) and $T_{C,Gd}$ (292 K). The common Curie temperature is induced by the strong exchange coupling which speeds up the demagnetization process in FeCo but slows down that in Gd. As shown in Fig.~\ref{fig5}(a), the Gd$_{21}$(FeCo)$_{79}$ shows FeCo dominant at all temperatures, whereas a transition from Gd to FeCo dominant is observed in the other sample. At low $T$, Gd dominates due to the larger magnetic moment. As $T$ increases, $\mathbf{m}_{net}$ reduces and vanishes at $T_{MC}$ = 75 K because of the faster demagnetization process in Gd. Above $T_{MC}$, $\mathbf{m}_{net}$ rises until a peak and then reduces to zero at $T_C$. Furthermore, we find that the magnetization dynamics near $T_{MC}$ [Fig.~\ref{fig5}(c)] is similar to the one at higher $T$ [Fig.~\ref{fig5}(b)], which can be understood by noticing the gradual change in effective fields such as $\mathbf{H}_A$ and $\mathbf{H}_I$. It is only at $T_{MC}$ that a sudden change occurs, and the effective fields diverge.

\begin{figure}
	\centering
	\includegraphics[width=\columnwidth]{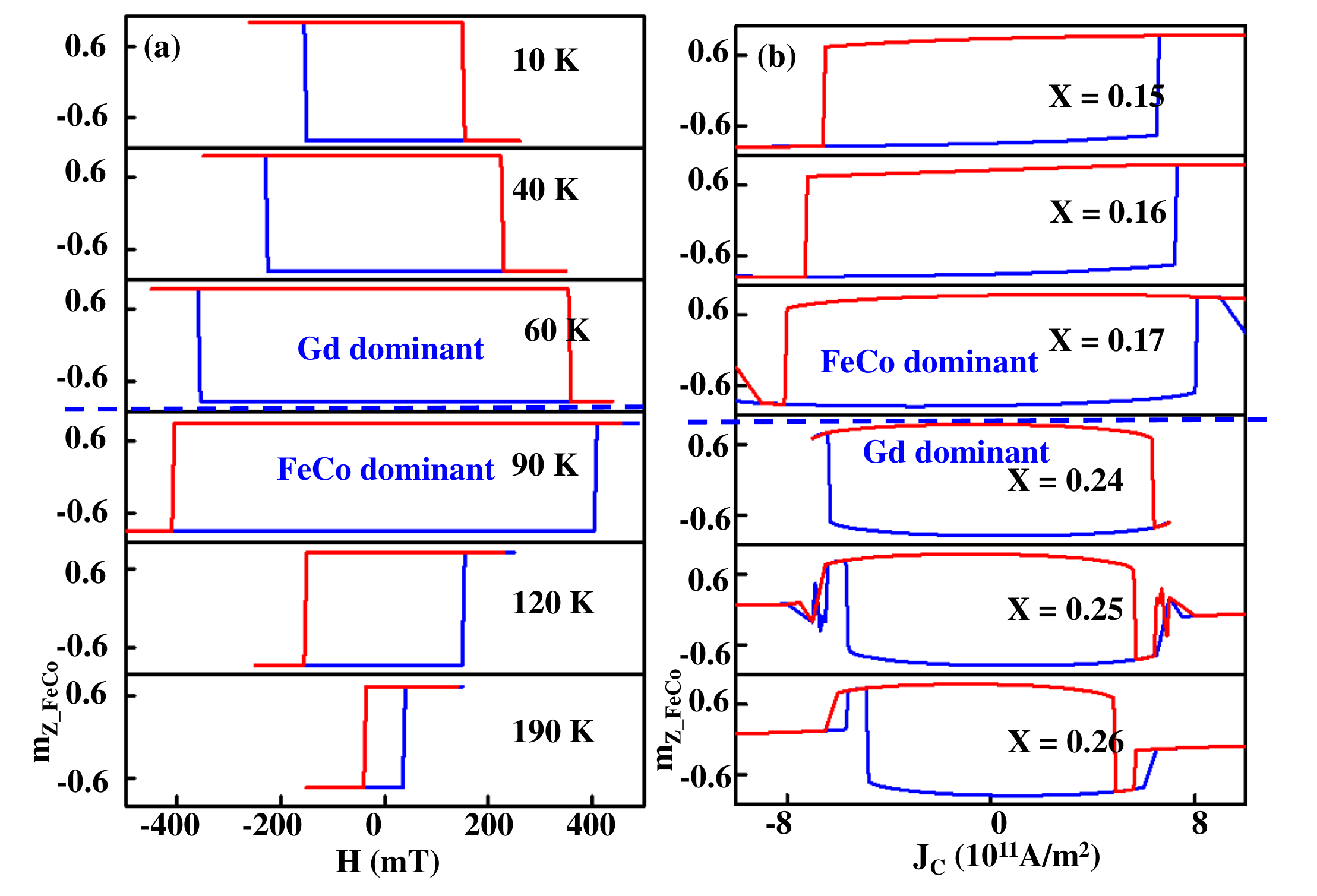}
	\caption{(a) $M$-$H$ loops at different $T$ in Gd$_{23}$(FeCo)$_{77}$ with $\mathbf{H}_X$ = 2 mT. The blue dot line denotes the transition from Gd to FeCo dominant. The switching-direction reversal and the peak in $H_C$ observed in experiments \cite{1882-0786-9-7-073001} are qualitatively reproduced. (b) $M$-$J_C$ loops at different $X$ with $\mathbf{H}_X$ = 2 mT. Three dynamic regions are identified, e.g., for $X$ = 0.25, successful switching happens for $5.7\times10^{11}$A/m$^2<J_C<6.4\times10^{11}$A/m$^2$, oscillation region for $6.4\times10^{11}$A/m$^2<J_C<8\times10^{11}$A/m$^2$, and distorted region for $J_C>8\times10^{11}$A/m$^2$ (i.e., $\mathbf{m}$ aligns to the $y$ direction).}
	\label{fig6}
\end{figure}

As shown in Fig.~\ref{fig6}(a), the competition between $\mathbf{m}_{FeCo}$ and $\mathbf{m}_{Gd}$ is also manifested in the $T$-dependent $M$-$H$ loops \cite{1882-0786-9-7-073001}. In addition to the reversal of switching direction, the $H_C$ reaches maximum at $T_{MC}$ to overcome the energy barrier ($E=-\mathbf{M}\cdot\mathbf{H}$). When $T$ is further increased (i.e., $T >T_{MC}$), both $\mathbf{m}_{FeCo}$ and $\mathbf{m}_{Gd}$ reduce, resulting in smaller exchange and anisotropy fields [Eq.~(\ref{eq3})] and hence a lower $H_C$. Furthermore, we find the $T_{MC}$ obtained from the $M$-$H$ loops is consistent with the equilibrium state calculation [Fig.~\ref{fig5}(a)], which is another evidence that the LLB model captures FiM dynamics. 

For practical reasons, $T$ is not preferred as the control parameter in device applications, whereas $X$ can be tuned during the deposition process. The change of $X$ shows similar results to that observed in the $T$ dependence. As $X$ is increased, the FiM changes from FeCo to Gd dominant, resulting in a reversal of both $M$-$H$ and $M$-$J$ loops \cite{PhysRevLett.118.167201,doi:10.1063/1.4962812}. Due to the vanishing $\mathbf{m}_{net}$ at $X_{MC}$, the spin torque diverges \cite{PhysRevLett.118.167201,PhysRevB.96.064406}. To show the capability of LLB model in capturing these effects, we have simulated the $X$-dependent current-induced switching at $T$ = 300 K. As shown in Fig.~\ref{fig6}(b), the switching direction reverses at $X_{MC}$ = 0.24 which separates FeCo and Gd dominant regions. In both regions, $\mathbf{m}_{net}$ is switched from down to up under positive current, indicating that the SOT-induced switching is determined by $\mathbf{m}_{net}$. This is different with the anomalous Hall effect (AHE), where $R_{AHE}$ is determined by $\mathbf{m}_{FeCo}$. In contrast to the magnetic-field-induced switching in Fig.~\ref{fig6}(a), no clear peak of critical switching current density ($J_{Crit}$) is observed, which is attributed to the increase of spin torque near $X_{MC}$. Interestingly, three dynamics regions are identified in our simulated $M$-$J$ loops. According to the subfigure of $X$ = 0.25 in Fig.~\ref{fig6}(b), $\mathbf{m}_{FeCo}$ is successfully switched from up to down for $5.7\times10^{11}$A/m$^2<J_C<$$6.4\times10^{11}$A/m$^2$. When the current exceeds $8\times10^{11}$A/m$^2$, the magnetization is aligned with $\mathbf{\sigma}$ due to the dominance of spin torque. The magnetization dynamics in these two regions are considered as typical behaviors which have also been observed in FM systems \cite{Cai2017}. For the $\mathbf{J}_C$ in between, however, an unexpected oscillation occurs. To understand this, we have simulated the SOT-induced switching in FM using the LLB model, which is realized by setting $X$ = 0, and no oscillation is observed. This indicates the important role of exchange coupling, which competes with $\mathbf{H}_I$ and $\mathbf{H}_X$, and the oscillation is obtained from the balance of all these interactions, which is a unique property in the system of perpendicular FiM switched by an in-plane current.

\section{CONCLUSION}
In conclusion, we have developed a theoretical model based on the modified LLB equation to systematically describe the effect of temperature, material composition, and spin torque in FiM. The effect of spin torque on the magnetization dynamics is studied in a FiM/HM bilayer, where the switching trajectory is found to be precession free due to the small precession field. Similar to the spin-torque switching in AFMs, the two sublattices are not always collinear, with a maximum angle deviation of 0.9 degree in our study. This cant between sublattices induces an oscillation region between the magnetization switching and the fully alignment with spin polarization, which is a unique behavior in the perpendicular FiM system. Our results of the spin-torque-induced magnetization dynamics can be helpful in understanding experimental results and in evaluating FiM-based devices. 

\begin{acknowledgments}
This work at the National University of Singapore was supported by CRP award no. NRF-CRP12-2013-01, NUS FRC R263000B52112, and MOE-2017-T2-2-114.
\end{acknowledgments}
\section*{Appendix A: Numerical simulation of FiM using the LLG model}
Before we study the LLB model, the LLG equation is widely used to qualitatively explain experimental findings in FiMs. As the FiM consists of antiferromagnetic coupled TM and RE sublattices, it is straightforward to apply one LLG equation to one sublattice \cite{OEZELT201528} as 
\begin{subequations}\label{eq10}
\begin{equation}
\begin{aligned}
\dot{\mathbf{M}}^a=&-\gamma^a\mathbf{M}^a\times(\mathbf{H}^a+h\mathbf{M}^b)+\alpha^a\mathbf{M}^a\times\dot{\mathbf{m}}^a\\
&-\gamma\mathbf{M}^a\times(\mathbf{M}^a\times\mathbf{H}^a_I),
\end{aligned}
\end{equation} 
\begin{equation}
\begin{aligned}
\dot{\mathbf{M}}^b=&-\gamma^b\mathbf{M}^b\times(\mathbf{H}^b+h\mathbf{M}^a)+\alpha^b\mathbf{M}^b\times\dot{\mathbf{m}}^b\\
&-\gamma\mathbf{M}^b\times(\mathbf{M}^b\times\mathbf{H}^b_I),
\end{aligned}
\end{equation} 
\end{subequations}
where the three terms on the right hand side are precession, Gilbert damping, and spin torque, respectively, and these two equations are coupled through the exchange terms, i.e., $h\mathbf{M}^b$ and $h\mathbf{M}^a$.

Despite the clear physical picture behind the coupled equations, the analytical study of FiM using these equations is complicated. Instead, another effective LLG model describing the net magnetization \cite{PhysRevLett.97.217202,PhysRevB.73.220402,PhysRevB.74.134404} is often used as  
\begin{equation}\label{eq11}
\begin{aligned}
\dot{\mathbf{M}}_{eff}=
&-|\gamma_{eff}|(\mathbf{M_{eff}}\times\mathbf{H^{eff}})+\alpha_{eff}(\mathbf{M_{eff}}\times\dot{\mathbf{M}}_\mathbf{eff})/M_{eff}\\
&-|\gamma_{eff}|(\mathbf{M_{eff}}\times(\mathbf{M_{eff}}\times\mathbf{H_I)}),
\end{aligned}
\end{equation} 
\begin{equation}\label{eq12}
\gamma_{eff}(T)=\frac{M_{RE}(T)-M_{TM}(T)}{M_{RE}(T)/|\gamma_{RE}|-M_{TM}(T)/|\gamma_{TM}|},
\end{equation}
\begin{equation}\label{eq13}
\alpha_{eff}(T)=\frac{\lambda_{RE}/|\gamma_{RE}|^2+\lambda_{TM}/|\gamma_{TM}|^2}{M_{RE}(T)/|\gamma_{RE}|-M_{TM}(T)/|\gamma_{TM}|},
\end{equation} 
where $\gamma_{eff}$ and $\alpha_{eff}$ are the effective gyromagnetic ratio and damping constant, respectively. By assuming a collinear TM and RE, we can demonstrate that the two LLG models are equivalent \cite{OEZELT201528}. The effective LLG model is firstly used to qualitatively explain the unexpected sign reversal of magnetoresistance (MR) in STT-switched CoGd layer \cite{PhysRevLett.97.217202}, where the coexistence of magnetization and angular-momentum compensation is observed for the first time. In addition, this model has been used in the ferromagnetic resonance (FMR) analysis \cite{PhysRevB.74.134404} and in explaining the ultrafast spin dynamics in GdFeCo \cite{PhysRevB.73.220402}. In addition, the quantitative study of the spin-torque-induced dynamics in FiM using the coupled LLG equations has been studied without considering the different g factors or damping constant between sublattices \cite{PhysRevLett.118.167201}. In this study, we choose $g_{FeCo}$ = 2.2, $g_{Gd}$ = 2 \cite{Kim2017}, $\alpha_{FeCo}$ = 0.01, $\alpha_{Gd}$ = 0.02, and perform numerical simulations of both magnetic-field and spin-torque switching using the effective LLG model, which is integrated via the fourth-order Runge-Kutta method \cite{GaraCervera2007NumericalM}. The simulation results are then compared with the experimental and the LLB results in the aspect of critical switching current and switching trajectory.

Limited by the fixed magnetization length in the LLG model, the simulations are performed at fixed $T$ which is less than $T_{C,Gd}$ due to the requirement of knowing both $M_{S,Gd}$ and $M_{S,FeCo}$ in Eq.~(\ref{eq11}). Therefore, we first choose $T$ = 200 K and study the field-induced switching with $t_{FiM}$ = 2 nm, and FiM radius $r_{FiM}$ = 25 nm. The $M_S$ of both sublattices are taken from their FM counterparts at $T$ = 200 K ($M_{S,Gd}=2.02\times10^6$ A/m \cite{PhysRev.132.1092} and $M_{S,FeCo}=1.73\times10^6$ A/m \cite{Crangle477}), resulting in a FeCo-dominant sample. As the critical switching field is independent on the damping constant, we are left with one free parameter, i.e., the crystalline anisotropy field ($H_K$). The magnitude of $H_K$ is determined by performing relaxation simulation, where the magnetization is required to return to the perpendicular direction, i.e., the magnetization is firstly tilted away from the $z$ direction (e.g., 70 degree tilt), and then, it is relaxed without any magnetic field or current. As a result, $H_K$ has to be larger than 0.9 T to maintain a perpendicular magnetization. Next, the effect of $H_K$ on $H_C$ is studied, where $H_C$ increases linearly with $H_K$. Consequently, the smallest $H_C$ = 200 mT is obtained using $H_K$ = 0.9 T, much larger than the experimental measured $H_C$ = 50 mT \cite{1882-0786-9-7-073001}. Furthermore, when the above procedures are repeated at other $T$ ($0 < T < 293 K$), we find that the sample is always FeCo dominant, failing to explain the experimental observed Gd-dominant region.
 
\begin{figure}
	\centering
	\includegraphics[width=\columnwidth]{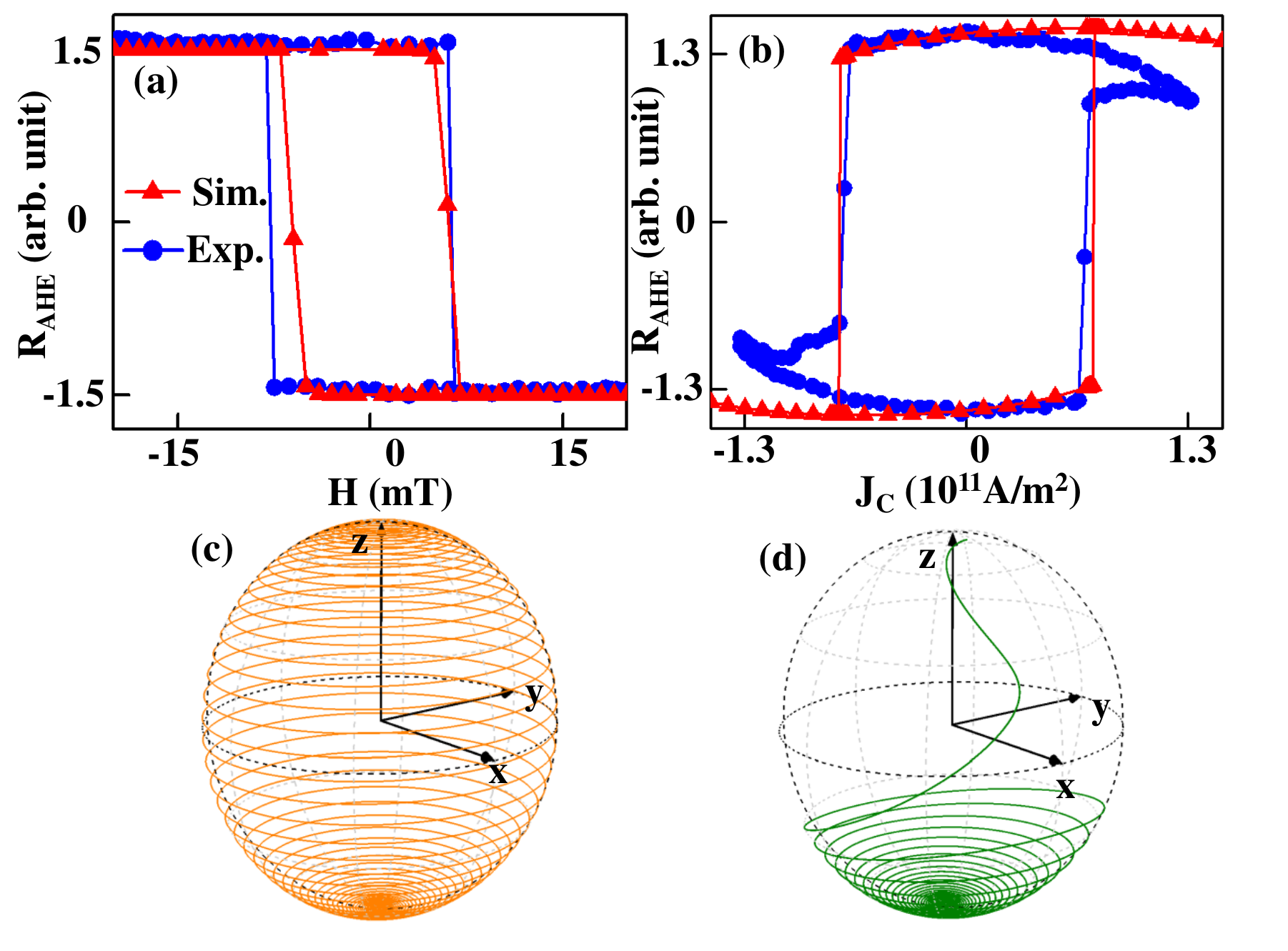}
	\caption{Experimental $R_{AHE}$ (blue circle) \cite{doi:10.1063/1.4962812} and simulated $m_{Z,FeCo}$ (red triangle) using the effective LLG model for (a) SOT-induced switching, and (b) magnetic-field switching with $M_{S,Gd}=3.2\times10^5$A/m and $M_{S,FeCo}=1.73\times10^6$A/m. (c) and (d) are the time evolutions of $\mathbf{m}_{FeCo}$ correspond to (a) and (b), respectively.}
	\label{fig7}
\end{figure}

However, the results from the LLG model can fit the experimental $M$-$H$ loop if the restrictions of taking $M_S$ from FM counterparts are removed [see Fig.~\ref{fig7}(a)], which can be justified by the difference in exchange coupling strength and lattice occupations between FiM and FM. We then simulate the current-induced switching using $M_S$ and $H_K$ obtained from the fitted $M$-$H$ loop, and a good fitting shown in Fig.~\ref{fig7}(b) is obtained with  $\alpha_{FeCo}$= 0.01, $\alpha_{Gd}$= 0.02 and spin polarization $P$ = 0.4.

Figs.~\ref{fig7}(c) and 7(d) show the trajectories of field and current switching, respectively. Both of them are similar to that in FMs \cite{Fukami2016}, which is expected since the effective LLG model treats the FiM as a single magnet. However, the current-induced magnetization dynamics predicted by the LLG model [see Fig.~\ref{fig7}(d)] is very different to that in the LLB model as shown in Fig.~\ref{fig3}(c), and more results from time-resolved experiments \cite{PhysRevB.75.064402} would be helpful to resolve this discrepancy. As a result, although the LLG model can reproduce the experimental $M$-$H$ and $M$-$J$ loops by fitting $M_S$, different sets of $M_S$ will be generated at different $T$, and some of them might be unrealistic. In addition, these $M_S$ are not correlated and cannot be explained using a unified theory. In contrast, the LLB model can reproduce the experimental results for all temperatures by using one set of parameters \cite{PhysRevB.97.184410}, which is attributed to its capability of capturing the $T$-dependent magnetization-length change. In this aspect, the LLB model is more suitable in studying FiM properties. 

\section*{Appendix B: Derivation of the LLB equation including spin torque}
The derivation of Eq.~(\ref{eq1}) starts from the lattice-site atomistic Landau-Lifshitz equation with an additional spin-torque term
\begin{equation}\label{eq14}
\mathbf{\dot{s}}=\gamma[\mathbf{s}\times(\mathbf{H+\mathbf{\zeta}})-\lambda\mathbf{s}{\times}(\mathbf{s}{\times}\mathbf{H})+(\mathbf{s}\times(\mathbf{s}\times\mathbf{H}_I))],
\end{equation}
\begin{equation}\label{eq15}
\mathbf{H}=\mathbf{H}_{ext}+(2D/\mu)s^z_i\mathbf{e}_z+\sum_{j\in{neig}}J_{ij}\mathbf{S}_{ij}/\mu,
\end{equation}
\begin{equation}\label{eq16}
<\zeta_a(t)\zeta_b(t')>=2{\lambda}T\delta_{ab}\delta(t-t')/(\gamma{\mu}_0),
\end{equation}
where $\mathbf{s}$ is the spin angular momentum, $\mathbf{\zeta}$ is the thermal field with the subscript representing different Cartesian components (i.e., $x$, $y$, and $z$), and $t$ is the time. The three terms on the right hand side of Eq.~(\ref{eq14}) represent precession, damping, and spin-torque effect, respectively. The exchange coupling in the last term of Eq.~(\ref{eq15}) only considers the influence of nearest neighbors, and Eq.~(\ref{eq16}) indicates that the sublattice spin is uncorrelated with respect to time and other Cartesian components. The direct simulation using Eq.~(\ref{eq14}) is known as atomistic modeling \cite{Radu2011,Ostler2012,PhysRevB.80.094418,PhysRevB.84.024407,doi:10.1002/9780470022184.hmm205}, and the information of magnetization dynamics is obtained by summing up all the lattice-site spins. Since the lattice constant is very small (a few angstroms), the atomistic model is limited to very small devices with diameter below 20 nm \cite{doi:10.1002/9780470022184.hmm205}. To simulate larger devices, a statistical model is developed based on Eq.~(\ref{eq14}), resulting in a single equation, i.e., Fokker Planck equation, which captures the spin dynamics as 
\begin{equation}\label{eq17}
\begin{aligned}
\frac{\partial{f}}{\partial{t}}&+\frac{\partial}{\partial(N)}\{\gamma\mathbf{N}\times\mathbf{H}-\gamma\mathbf{N}\times(\mathbf{N}\times(\lambda\mathbf{H}+\mathbf{H}_I))\\
&+\frac{\gamma{\lambda}T}{\mu_0}[\mathbf{N}\times(\mathbf{N}\times\frac{\partial}{{\partial}N})]\}f=0,
\end{aligned}
\end{equation} 
where $f$ is the spin-distribution function, and $\mathbf{N}$ is a vector on a sphere with $|\mathbf{N}|$ = 1. Then, the spins are transformed to magnetization through 
\begin{equation}\label{eq18}
\mathbf{m}\equiv<\mathbf{s}>={\int}d^3N\mathbf{N}f(\mathbf{N},t),
\end{equation}
and Eq.~(\ref{eq17}) becomes
\begin{equation}\label{eq19}
\dot{\mathbf{m}}=\gamma[\mathbf{m}\times\mathbf{H}]-\Lambda_N\mathbf{m}-\gamma\lambda<\mathbf{s}\times[\mathbf{s}\times\mathbf{H}]>.
\end{equation}  
However, Eq.~(\ref{eq19}) is difficult to solve due to the mixture of $\mathbf{m}$ and $\mathbf{s}$, which can be resolved by applying the mean field approximation (MFA) \cite{PhysRevB.86.104414,PhysRevB.84.024407}, resulting in an explicit equation showing as Eq.~(\ref{eq1}). This process of model development is summarized as a flowchart in Fig.~\ref{fig1}(b).

\end{document}